\documentclass[aps,twocolumn,nofootinbib,superscriptaddress,amsmath,amssymb]{revtex4-1}
\newcommand{\bd}{{\bf d}}

\newcommand{\bpartial}{\boldsymbol{\partial}}
\newcommand{\rd}{{\mathrm d}}

\newcommand{\bnabla}{\boldsymbol{\nabla}}
\font\bigastfont=cmr10 scaled \magstep 1
\def\bdot{\hbox{\bigastfont .}}
\newcommand{\CR}{\mathcal R}
\newcommand{\CF}{\mathcal F}
\newcommand{\beq}{\begin{equation}}
\newcommand{\eeq}{\end{equation}}
\usepackage{bm}
\usepackage{xcolor}
\usepackage{hyperref}
\hypersetup{
    colorlinks=true,
    citecolor=magenta,
    linkcolor=violet,
    filecolor=cyan,      
    urlcolor=magenta,
}
\definecolor{MyDarkRed}{rgb}{0.7,0,0}

\begin{document}
\title{Direct correspondence between Newtonian gravitation and general relativity}
\author{Thomas Buchert}
\affiliation{Univ Lyon, Ens de Lyon, Univ Lyon1, CNRS, Centre de Recherche Astrophysique de Lyon UMR5574, F-69007, Lyon, France\\ Email: buchert@ens-lyon.fr}
%
%
\begin{abstract}
We present a strategy to obtain equations of general relativity for an irrotational dust continuum within a flow-orthogonal foliation of spacetime from the equations of Newtonian gravitation, and vice versa, without employing a weak field expansion or a limiting process on the speed of light. We argue that writing Newton's equations in a Lagrangian frame and relaxing integrability of vector gradients is sufficient to obtain equations that are identical to Einstein's equations in 
(3+1)-form when respecting the Lorentzian signature of the time parametrization. 
We discuss implications and provide an outlook on how to extend the obtained correspondence to more general spacetimes.
\end{abstract}
\maketitle
\section{Newtonian limits and weak fields:\\ is this subject settled ?}
\label{intro}
The question of how to obtain the Newtonian limit of general relativity enjoys various answers. Many of the practical implementations of a Newtonian limit are heuristic by e.g. expanding Einstein's equations with respect to a flat background spacetime and keeping only linear terms as perturbations of metric components (weak field approximation) together with sending the causality constant $1/c$ to zero, thus ``opening up'' the local light cones. A systematic approach is the \textit{frame theory} by Ehlers \cite{ehlers:frametheory} that comprises both theories in a single theory using the causality constant as a parameter, built on earlier efforts around the Newton-Cartan theory (for details and references see Ehlers' work \cite{ehlers:frametheory}, the editorial to Ehlers' work \cite{bm}, and Ehlers' investigation of examples \cite{ehlers:examples}). Ellis \cite{ellis:relativistic} has nicely compared the (1+3)-Einstein equations and their correspondences in Newtonian gravitation. 
There are also proposals of a dictionary between the theories for metric perturbations to linear order \cite{N-GRnumerics1,wald:dictionary} that have been applied in the context of general-relativistic exact solutions and numerical simulations  \cite{KoksbangHannestad}, \cite{N-GRnumerics2}. All of these approaches imply that both theories are substantially different and that one has to neglect terms in Einstein's equations to obtain back the Newtonian equations of gravitation. It is, however, a matter of definition of the Newtonian limit and the way the Newtonian equations of gravitation are written. 

In this Letter, we will investigate a two-step strategy for arriving at Einstein's field equations from Newton's equations, in the context of an irrotational dust matter model, that does not need elements of an approximation. \textit{As a first step}, we will demonstrate that the Newtonian equations and Einstein's equations in a flow-orthogonal foliation of spacetime are algebraically correspondent, if the former are written in the Lagrangian rest frame of the dust. \textit{As a second step}, we will show that the Newtonian equations become identical to their relativistic counterparts, if an integrability requirement of the resulting tensor coefficients is relaxed in the Newtonian equations. These statements extend to variables such as connection, Ricci and Weyl curvatures, defined geometrically within general relativity but are also algebraically present in the Newtonian equations, if written in the Lagrangian frame. 

We organize this Letter as follows. Sec.~\ref{newton} recalls Newton's equations for a self-gravitating continuum of dust.
It is then argued in Sec.~\ref{recipee} that the Lagrangian form of Newton's equations can be subjected to a ``recipe'' that follows an idea by Einstein to ``throw out the Euclidean embedding space'' and that employs Cartan coframes defining nonintegrable deformations of the fluid. We so obtain Einstein's equations in a flow-orthogonal (3+1)-setting.  Sec.~\ref{discussion} discusses the result and provides an outlook on how to generalize the obtained correspondence. 

\vspace{5pt}

\noindent
\textit{NOTATIONS:} Bold notation will be used for vectors and forms. The vector product is denoted by $\times$, while the wedge product $\wedge$ denotes the antisymmetrized tensor product $\otimes$, with components $({\bm a} \otimes {\bm b})_{ij} = a_i b_j$. We define symmetrization and antisymmetrization of indices by $v_{(i,j)} = \tfrac{1}{2}(v_{i,j}+v_{j,i})$, $v_{[i,j]} = \tfrac{1}{2}(v_{i,j}-v_{j,i})$, respectively; $i,j,k \ldots  = 1,2,3$. Spatial derivatives with respect to Eulerian coordinates $\bm{x}$ are denoted by a comma, while later spatial derivatives with respect to Lagrangian coordinates $\bm{X}$ are denoted by a vertical slash; an overdot will be used to represent the Lagrangian time derivative; summation over repeated indices is understood.

\section{Newtonian equations for a dust continuum}
\label{newton}
We will employ a hydrodynamic picture and consider a self-gravitating continuum of dust, i.e. pressureless matter with rest mass density $\varrho ({\bm x},t)$ and velocity field ${\bm v}({\bm x},t)$ (the words `matter', `dust', `fluid' will be used interchangeably). Both fields are represented in terms of Eulerian (inertial, nonrotating) coordinates $\bm x$ and a time parameter $t$. The acceleration field is equivalent to the gravitational field strength $\bm{g}({\bm x},t)$ due to the equivalence of inertial and gravitational mass,
\begin{equation}
\label{euler}
    \frac{\rd}{\rd t}\bm{v}=\bm{g} \quad ; \quad  \frac{\rd}{\rd t} :=\frac{\partial}{\partial t}\Big\vert_ {\bm{x}}  + \bm{v}\cdot\bnabla = \frac{\partial}{\partial t}\Big\vert_ {\bm{X}}\ ,
    \end{equation}
where we introduced the Lagrangian time derivative $\rd / \rd t$ that reduces to a partial time derivative in a Lagrangian coordinate system $\bm X$. 
In Eulerian coordinate components, we write Euler's equation \eqref{euler} and its Eulerian spatial derivative:
\begin{equation}
\label{eulerderivative}
    \frac{\partial}{\partial t} v^{i}+v^{k} v^i_{\ , k}=g^{i} \quad , \quad  \frac{\mathrm d}{{\mathrm d} t} v^i_{\ , j}+v^i_{\ , k} v^k_{\ , j} = g^i_{\ , j}\ .
    \end{equation}
The Newtonian continuum theory of gravitation is a vector theory, so that the sources of the curl and divergence of  $\bm{g}$ suffice to define the complete set of field equations (up to a harmonic vector field): 
\beq
\label{fieldequations}
 \bnabla \times \bm{g}=\bm{0}  \quad;\quad  \bnabla\cdot\bm{g}=\Lambda-4 \pi G \varrho \ ,
\eeq
where $G$ denotes Newton's gravitational coupling constant; for completeness we included the cosmological constant $\Lambda$ (here with dimension ${\rm time}^{-2}$). The rest mass density $\varrho$ obeys the continuity equation,
\beq
\label{continuity}
\frac{\rd}{\rd t} \varrho+\varrho \bnabla\cdot\bm{v}=0 \ .
\eeq
The Euler-Newton system comprises \eqref{euler}, \eqref{fieldequations} and \eqref{continuity}. 
This overdetermined set of equations can be written as a set of five equations for five variables ($\varrho, v^i, \Phi$) through introduction of the gravitational potential, $\bm{g} = :-\bnabla \Phi$. 

In this Letter we will restrict ourselves to irrotational flows, $v_{[i,j]} = 0$.
For later considerations we add the Newtonian \textit{gravitoelectric tidal tensor} (inserting \eqref{eulerderivative} and \eqref{fieldequations} in the second line):
\beq
\begin{aligned}
\label{tidalelectric}
{\mathcal E}_{i j} :&=  g_{(i ,j)} -\frac{1}{3} \delta_{ij}  g^k_{\ , k} \\
&= \frac{\mathrm d}{{\mathrm d} t} v_{(i, j)}+ v^k_{\ , (i}v_{j) , k} - \frac{1}{3} \delta_{ij} (\Lambda - 4\pi G\varrho)\ .
\end{aligned}
\eeq

\section{Realizing Einstein's vision of a transition strategy}
\label{recipee}

In his Kyoto address in December 1922 \cite{ishiwara}, Albert Einstein expressed an intuition of an interesting
strategy that we will freely interpret in this Letter. He said:
\textit{If all accelerated systems are equivalent, then Euclidean geometry cannot hold in all of them. To throw out geometry [the Euclidean embedding space]
and keep [vectorial] physical laws is equivalent to describing thoughts without words.
We must search for words before we can express thoughts [...]
This point remained unsoluble to me until 1912, when I suddenly realized,
that Gau{\ss}'s theory of surfaces holds the key for unlocking this mystery. [...]}.

\vspace{3pt}

We have interpreted Einstein's words [in brackets] to infer that the writing of Newton's equations in terms of vectors 
as functions of inertial coordinates implies an embedding into a global vector space.
A modern strategy to ``throw out'' this embedding space consists in (i) moving to a Lagrangian representation of Newton's equations, introducing (intrinsic and noninertial) Lagrangian coordinates, and (ii) relaxing integrability of the resulting tensor coefficients, i.e. replacing exact gradients by general one-form fields.

To achieve (i), we introduce a one-parameter family of spatial diffeomorphisms to Lagrangian coordinates $\bm{X}$, labeling fluid parcels along their trajectories, identified with their Eulerian positions at some initial time $t_{\rm i}$ \cite{ehlersbuchert}:\footnote{Henceforth, we use the indices $a,b,c \ldots$ as counters, while $i,j,k \ldots$ remain coordinate indices referring to an exact basis. We realize that the vector components $f^a$ are to be written with counters, but if there exists an embedding space endowed with coordinates $x^i = f^{i\equiv a}$, then these counters are also coordinate indices.}
\beq
\label{diffeo}
(\bm{X},t)\mapsto (\bm{x},t) = ({\bm f}({\bm{X},t}),t) \ ; \ \bm{X}:= \bm{f} (\bm{X}, t_{\rm i}) \ .
\eeq
We call $\bm{f}$ the field of trajectories and ${\bf d} f^a = f^a_{\ |k}{\bf d} X^k$ the \textit{Lagrangian deformation gradient}, represented in the exact Lagrangian basis. 
Lagrangian coordinates are intrinsic to the fluid continuum and they can as well be used as coordinates in a local chart around a point in a spatial Riemannian manifold. Let us consider the following bilinear metric form, the first representation of which we call the \textit{Lagrangian metric},
\beq
\label{metricE}
{^3}{\bm\delta} : =\delta_{ab}\,f^a_{\;|i} f^b_{\;|j} \bd X^i \otimes \bd X^j = \delta_{ij} \;\bd x^i \otimes \bd x^j \ .
\eeq
This metric is Euclidean, since we can find a one-parameter family of diffeomorphisms, namely ${\bm X} = {\bm h}({\bm x},t)$, with ${\bm h} = {\bm f}^{-1}$, that transforms the first representation into the second. 

A Riemannian 3-metric ${^3}{\bm s}$ can be written in terms of three spatial \textit{Cartan coframes} ${\bm\eta}^a$ that provide a more elementary ``reference body'' than the metric, 
\beq
\label{metricR}
{^3}{\bm s}:=\delta_{ab}\;\boldsymbol{\eta}^a \otimes  \boldsymbol{\eta}^b = 
\delta_{ab} \;\eta^a_{\;i}\,\eta^b_{\;j} \,\bd X^i \otimes \bd X^j \ . 
\eeq
We notice that the mapping
\beq
\label{restriction}
\boldsymbol{\eta}^a \; \mapsto \; \bd f^a \ ,
\eeq
which we henceforth call \textit{Euclidean restriction},
implies that the Riemannian 3-metric reduces to 
the Euclidean metric in the Lagrangian representation. 
In the integrable case of exact one-form fields $\bd f^a$, the Lagrangian metric embodies the 
geometry of the fluid, however, still embedded into Euclidean space.

We henceforth use the wording \textit{integrable} when we want to express that the coefficient matrix $\eta^a_{\ i}$ of $\boldsymbol{\eta}^a =\eta^a_{\ i} \bd X^i$ 
can be obtained through spatial derivatives of vector components $f^a$, i.e. $\boldsymbol{\eta}^a$ are exact, and we say \textit{generalized gradient} when we mean relaxation of \textit{integrability} that realizes step (ii) of the outlined strategy.

In what follows we will apply the outlined two-step strategy, where we first concentrate on kinematic properties of the fluid. To this end we are going to find the analogy to the Newtonian 
\textit{velocity gradient} $v^a_{\ ,b}$, now written both with counter indices, since there will no longer be a reference to an exact Eulerian basis after relaxing integrability in the sense of inverting \eqref{restriction}. Relating this to the Lagrangian gradient of $\bm v = \dot{\bm f}({\bm X},t)$ involves the inverse transformation ${\bm h}({\bm x},t)$: $v^a_{\ ,b}=v^a_{\ |k}h^k_{\ ,b} = {\dot f}^a_{\ |k}h^k_{\ ,b}$. 

Moving to the nonintegrable form of the velocity gradient, we have to introduce the inverse 
matrix to  ${\eta}^a_{\ k}$. We define three {\it frame fields}, ${\bm e}_b$
(the {\it Dreibein} at the worldlines of fluid parcels), being dual to Cartan's coframe 
fields. We express both in the respective local basis systems ($\bd X^k$ for forms and $\bpartial_{X^k}$ for vectors):
\begin{equation}
\label{framefields}
\boldsymbol{\eta}^a = {\eta}^a_{\ k}\,\bd X^k \,;\,{\bm e}_b =e_b^{\ k}\,\bpartial_{X^k}
\,;\,{\eta}^a_{\ k} e_b^{\ k} = \delta^a_{\ b}\,;\,{\eta}^a_{\ k} e_a^{\ \ell} = \delta_k^{\ \ell}\; .
\end{equation}
Consequently, the nonintegrable form of the Newtonian velocity gradient is represented by
\beq
\label{analogyvelocitygradient1}
\Theta^a_{\ b}:= {\dot\eta}^a_{\ k} e_b^{\ k}\ .
\end{equation}
It is expressed in the nonexact basis (remember that the velocity gradient has both
Eulerian values and derivatives with respect to Eulerian coordinates).
We transform this object into our local exact (Lagrangian) basis with the help of the transformation
matrices (\ref{framefields}) and arrive at:
\begin{subequations}
\beq
\label{analogvelocitygradient2}
\Theta^i_{\ j}:= e_a^{\ i}\eta^b_{\ j}\,\Theta^a_{\ b}=e_a^{\ i}{\dot\eta}^a_{\ j}\ . 
\eeq
This field can be entirely expressed in terms of coframe fields through the algebraic identity
\beq
\label{frametransformation}
e_a^{\ i} = \frac{1}{2J}\;\epsilon_{abc}\epsilon^{ik\ell}\,\eta^b_{\;k}\,\eta^c_{\;\ell}\ ,
\eeq
with the Levi-Civit\`a symbol $\epsilon_{abc}$, and the nonintegrable analog of the Jacobian of the spatial diffeomorphism \eqref{diffeo},
\beq
\label{Jacobian_art}
J: =\det (\eta^a_{\ i}) = \frac{1}{6}
\epsilon_{abc}\epsilon^{ijk}\,\eta^a_{\ i}\eta^b_{\ j}\eta^c_{\ k}\ .
\eeq
\end{subequations}
We notice that the variable \eqref{analogvelocitygradient2} that generalizes the Newtonian velocity gradient has mixed indices, which holds true for the transformation of other Newtonian fields too. 
The expansion tensor is then formed by lowering the upper index using the nonintegrable form of the Lagrangian metric \eqref{metricR}, $\Theta_{ij} = \delta_{ab}\eta^a_{\ i}\eta^b_{\ k}\Theta^k_{\ j} = \delta_{ab}\,\eta^a_{\ i}{\dot\eta}^b_{\ j}$, with the rate of expansion $\Theta^k_{\ k} = :\Theta$. The vanishing of the vorticity, $v_{[i,j]}=0$, so translates to the symmetry condition $\Theta_{[ij]}=0$.\\

Turning now to dynamical properties of the dust fluid, we introduce the nonintegrable form of the field strength gradient along the above lines, $g^a_{\ ,b} \mapsto \CF^a_{\ b}={\ddot\eta}^a_{\ k} e_b^{\ k}$, 
${\CF}^{i}_{\ j}:= e_a^{\ i}\eta^b_{\ j}\,\CF^a_{\ b}=e_a^{\ i}{\ddot\eta}^a_{\ j}$ and $\CF_{ij} = \delta_{ab}\,\eta^a_{\ i}{\ddot\eta}^b_{\ j}$,
yielding the nonintegrable version of Euler's equation \eqref{eulerderivative}: 
\begin{subequations}
\label{Fequations}
\beq
\label{generalizedeuler}
{\CF}^{i}_{\ j} = \dot{\Theta}^i_{\ j} + \Theta^i_{\ k}\Theta^k_{\ j} \ .
\eeq
The field equations \eqref{fieldequations} generalize to the set
\beq
\label{generalizedfieldequations}
{\CF}^k_{\ k} = \Lambda - 4\pi G \varrho \quad;\quad
{\CF}_{[ij]} = 0 \ .
\eeq
In the integrable version of Newton's equations, these are enough to determine the gravitational field. However, in the nonintegrable version the tracefree symmetric part must be part of the gravitational field tensor. We therefore add the nonintegrable form of the Newtonian tidal tensor \eqref{tidalelectric} (omitting here the redundant symmetrization):
\beq\begin{aligned}
\label{electricweyl}
- {E}^i_{\ j} :&= \CF^i_{\ j} - \frac{1}{3}\CF^k_{\ k}\delta^i_{\ j} \\
&= \dot{\Theta}^i_{\ j} + \Theta^i_{\ k}\Theta^k_{\ j} - \frac{1}{3} (\Lambda - 4\pi G \varrho)\delta^i_{\ j}  \ ,
\end{aligned}\eeq
\end{subequations}
where we introduced a sign convention that we will explain below. 

We now show that Equations~\eqref{Fequations}, together with the nonintegrable Lagrangian form of \eqref{continuity}, 
$\dot\varrho + \Theta \varrho = 0$, are identical to Einstein's equations in a fluid-orthogonal foliation of spacetime \textit{via} the \textit{definition} of a new (from the Newtonian point of view auxiliary) field:
\beq
\label{Ricci}
- \CR^i_{\ j}:={\dot\Theta}^i_{\ j} +\Theta {\Theta}^i_{\ j} - (\Lambda + 4\pi G\varrho)\delta^i_{\ j}  \ . 
\eeq
Equation~\eqref{Ricci} implies a key equation of the correspondence: with \eqref{generalizedeuler} we obtain a relation of the generalized field strength gradient to this newly defined field:
\beq
\label{key1}
{\CF}^{i}_{\ j} = - \CR^i_{\ j} + (\Lambda + 4\pi G\varrho)\delta^i_{\ j} +
 \Theta^i_{\ k}\Theta^k_{\ j} - \Theta \Theta^i_{\ j} \ .
\eeq
In the geometrical context of general relativity this field is the spatial Ricci tensor, $\CR_{ij} = \delta_{ab}\eta^a_{\ i}\eta^b_{\ k}\CR^k_{\ j}$, the key equation \eqref{key1} is known to emerge from the Gau{\ss} embedding equation using the nonintegrable Euler equation \eqref{generalizedeuler}:
the components of the generalized Newtonian field strength gradient form components of the spacetime Riemann tensor, 
$-{\CF}^{i}_{\ j} = {^4}R^i_{\ 0 j 0}$. 
Imposing the field equations \eqref{generalizedfieldequations}, the trace of \eqref{key1} becomes the energy constraint, and the antisymmetric part of \eqref{key1} vanishes due to the vanishing of the vorticity in a flow-orthogonal foliation: ${\CF}_{[ij]} = \delta_{ab}\,\eta^a_{\ [i}{\ddot\eta}^b_{\ j]} =  (\delta_{ab}\,\eta^a_{\ [i}{\dot\eta}^b_{\ j]})^{\bdot} = 0$. 
The nonintegrable form of the Newtonian tidal tensor \eqref{electricweyl} is the spatially projected gravitoelectric part of the Weyl tensor \cite{ehlersbuchert:weyl}. 
It reduces to \eqref{tidalelectric} in the Euclidean restriction \eqref{restriction}, up to a sign convention.\footnote{The sign convention difference arises, since in the geometrical context we consider $E_{ij}$ as (``passive'') curvature, while in Newtonian theory the corresponding field is defined ``actively'' in terms of gravitational acceleration. This remark also applies to the extrinsic curvature $K_{ij}$ vs. expansion $\Theta_{ij}= -K_{ij}$.}

For completeness we list the Einstein equations for an irrotational dust fluid in a flow-orthogonal foliation of spacetime in the usual (3+1)-representation:\footnote{The metric signature is taken to be $(-,+,+,+)$, and the speed of light $c=1$.
Greek indices run through $\mu , \nu \ldots = 0,1,2,3$, 
and the semicolon denotes covariant derivative with respect to the 4-metric, while a double vertical slash denotes covariant spatial derivative with respect to the 3-metric ${^3}\bm s$ with components $s_{ij}$.}
\begin{subequations}
\begin{eqnarray}
&{\dot \varrho} + \Theta\varrho = 0 \ ;\\
&{\dot s}_{ij} = 2\,s_{ik} \Theta^k_{\ j}\ \ ; \ \ \Theta_{[ij]} = 0 \  ;\\
&{\dot\Theta}^i_{\ j} +\Theta {\Theta}^i_{\ j} = - \CR^i_{~j}+(\Lambda + 4\pi G\varrho)\delta^i_{\ j} \ ;\\
&\Theta^2 - \Theta^i_{\ j}\,\Theta^j_{\ i} = - {\CR}^k_{\ k} + 2\Lambda +16\pi G \varrho \ ; \\
&  \Theta^k_{\ j ||k} - \Theta_{|| j}\,=\,0\ . \label{momentumconstraints}
\end{eqnarray}\label{admdust2}
\end{subequations} 
\noindent The first equation arises from $T^{\mu\nu} = \varrho u^\mu u^\nu$, 
with the conservation law $T^{\mu\nu}_{\ \ ; \nu} = 0$, while the second defines the expansion tensor (or minus the extrinsic curvature); the third are its $6$ evolution equations that are identical to the nonintegrable Euler equation \eqref{generalizedeuler} by redefining $\CR^i_{\ j}$ through \eqref{key1}, and the fourth is one of the four constraint equations,  the energy constraint, all -- as shown -- arising from our strategy. 

The momentum constraints \eqref{momentumconstraints} seem to not directly arise from the Newtonian system because their Euclidean restriction \eqref{restriction} does not imply a constraint. This can be traced back to the fact that the spatially projected gravitomagnetic part of the Weyl tensor $H_{ij}$ vanishes in the Euclidean restriction \eqref{restriction}, in the current setting: 
\beq
\label{magneticweyl}
-H^i_{\ j} =\frac{1}{J} \epsilon^{ik\ell} \Theta_{jk || \ell}\ \mapsto \ 0\ ; \ J \ne 0 \ ;
\eeq
$H_{[ij]}=0$ implies \eqref{momentumconstraints}. This result is in agreement with the Newtonian limit in Ehlers' frame theory \cite{ehlersbuchert:weyl}, and here trivially follows \textit{via} integrability that implies the commutation of second derivatives, see subsections III.A.3 and III.A.4 in \cite{rza1}. 

We can derive \eqref{momentumconstraints} by starting with the trivial Newtonian identity that second derivatives of the velocity commute, $v^a_{\ , b\,c}-v^a_{\ ,c\,b}=0$. Calculating $v^a_{\ ,b} = v^i_{\ | j}f^a_{\ |i}h^j_{\ ,b}$, transforming the second derivatives to Lagrangian coordinates and projecting onto the Lagrangian basis (step (i) of our strategy) yields:\footnote{Notice that the (noncovariant) Christoffel connection coefficients $\Gamma^j_{\ n \ell}$ do not vanish in the Euclidean restriction \eqref{restriction}: they result in a Newtonian integrable connection (since the Lagrangian coordinate system is noninertial):
\beq
\Gamma^j_{\ \ell n}=\Gamma^j_{\ n \ell} \mapsto {}^{\rm N}\Gamma^j_{\ n \ell} = h_{\ ,c}^{j} f^c_{\ | \ell n} = - h^j_{\ ,ab}f^a_{\ | \ell}f^b_{\ | n}={}^{\rm N}\Gamma^j_{\ \ell n}
\eeq
(both forms appear in the calculation of \eqref{mom}).
However, the Euclidean restriction of the (covariant) Cartan connection $\bd{\bm\eta}^a = :-{\bm\omega}^a_{\ b} \wedge {\bm\eta}^b$ vanishes, i.e. the covariant requirement of integrability, ${\bf d}^2  f^a = {\bf 0}$, with ${\bf d}^2 := {\bf d} \circ {\bf d}$, holds.\\
Notice also that, starting with the transformation of the vector, $v^a = f^a_{\ | i}v^i$, instead of its gradient, will result in extra terms proportional to the Lagrangian velocity $v^i = 0$, which consequently vanish in a Lagrangian frame.}
\begin{eqnarray}
\label{mom}
&&(v^a_{\ , bc}-v^a_{\ , cb}) (h^k_{\ ,a}f^b_{\ | n}f^c_{\ |\ell}) = v^k_{\ | n\ell}-v^k_{\ |\ell n}\nonumber\\
&&+ {}^N \Gamma^j_{\ \ell n} v^k_{\ | j} - {}^N \Gamma^j_{\ n\ell} v^k_{\ | j}
+ {}^N \Gamma^k_{\ \ell i} v^i_{\ | n} - {}^N \Gamma^k_{\ n i} v^i_{\ | \ell}\nonumber\\
&&= v^k_{\ | n || \ell}-v^k_{\ | \ell || n} = \epsilon_{i n\ell}\epsilon^{ijm} v^k_{\ | j || m} \ .
\end{eqnarray}
Relaxing integrability (step (ii) of our strategy) then results in the Peterson-Mainardi-Codazzi identity in the flow-orthogonal foliation [Sect.8.3]\cite{alcubierreGR}, \cite{eric} (using \eqref{magneticweyl} in the second equality):
\begin{equation}
\label{mag}
\Theta^k_{\ n || \ell} - \Theta^k_{\ \ell || n} = \epsilon_{i n\ell}\epsilon^{ijm} \Theta^k_{\ j || m} = -J \epsilon_{i n \ell} H^{ik} \  ,
\end{equation}
the trace of which ($k=\ell$) is \eqref{momentumconstraints}. Note that the trace of the gravitomagnetic part of the Weyl tensor vanishes due to the symmetry condition $\Theta_{[pj]} = 0$. In the integrable case both sides are identically zero.\\

An important remark is in order here. Both steps (i) and (ii) are crucial for our transformation strategy, but notice that step (ii) produces a nonintegrable deformation leading to a general description of spatial deformations in terms of Cartan coframe fields. Therefore, following from step (ii), we can derive all the elements of geometry like a nonintegrable connection and curvature via Cartan's structure equations in space, $\bd {\bm\eta}^a = -\bm\omega^a_{\ b}\wedge \bm\eta^b \ne {\bf 0}$ and $\bm\Omega^a_{\ b} := \bd \bm\omega^a_{\ b} + \bm\omega^a_{\ c}\wedge\bm\omega^c_{\ b}\ne {\bf 0}$, together with the spatial Bianchi identities $\bd^2  {\bm\eta}^a = {\bm 0}$ and $\bd^2  {\bm\omega}^a_{\ b} = {\bm 0}$.\\

The final element of the correspondence arises when constructing the spacetime metric. We notice that the Newtonian equations and the (3+1)-Einstein equations appear to be parametrized by the coordinate $t$: our two-step strategy produces the correct equations. However, when constructing a 4-dimensional spacetime, the introduction of the Lorentzian signature in the 4-metric is required: ${^4}{\bm s}:=- \bd t \otimes \bd t + \delta_{ab} \;\eta^a_{\;i}\,\eta^b_{\;j} \,\bd X^i \otimes \bd X^j$.
The Euclidean restriction \eqref{restriction} is then extended to spacetime and becomes the restriction to Minkowski spacetime. We understand that the Lagrangian representation is a crucial cornerstone of the correspondence: Lagrangian observers are at rest and ``do not see" the local light cone, since they do not experience a boost. The Lagrangian observers have just to be told that their distances
in time direction count negatively in a causal 4-dimensional spacetime.

\section{Summary and discussion}
\label{discussion}

We looked at the Newtonian equations for self-gravitating systems in the Lagrangian frame together with a generalization to a nonintegrable form of the Newtonian deformation gradient ${\bf d} f^a$. 
We restricted our investigation to the matter model of irrotational dust. We argued that the nonintegrable form is equivalent to Einstein's equations in a flow-orthogonal (3+1)-foliation of spacetime. We observe that
there is no weak field approximation and no limiting process to be performed. None of the parts in Einstein's equations are neglected, which paints an alternative picture to the current understanding of Newtonian and post-Newtonian dynamics. Newton's theory appears to be stronger than believed by employing a modern interpretation.
It will be interesting to revisit predictions of general relativity where 
Newtonian predictions in Eulerian representation appear to fall short.

We could furthermore argue that the integrable (Lagrange-Newton) form is a measure-zero representation of the general form that, from a pragmatic point of view, can never be realized:
any realization, e.g. a numerical implementation of an exact gradient ${\bf d} f^a$, will be limited by finite precision and generically produces a nonintegrable field. 
Newtonian gradients form a measure zero set of fluid deformations and meet the strong condition ${\bf d}^2 f^a = {\bf 0}$ (second derivatives commute). Hence, any realization will instantly produce the nonintegrable form, and therefore a nonintegrable connection and curvature via Cartan's structure equations. A small perturbation of strict integrability can lead to strong curvature without there being a smooth limit to Euclidean space.
As a pronounced summary: an attempt to realize Newtonian dynamics in the Lagrangian representation will, in practice, be a realization of general relativity.

Mathematically, the requirement of integrability has nevertheless interesting and important implications. We think of the spatial integration of an integrable field vs. a nonintegrable field.\footnote{The difference can be best seen by performing a Hodge decomposition of Cartan forms into exact, co-exact and harmonic forms, for a component writing of one-forms in this context see \cite{rza4}.} Integrable parts will allow for a transformation to surface integrals on the boundary of a spatial domain, while nonintegrable parts remain in the bulk. An example is the backreaction problem in cosmology, i.e. the impact of inhomogeneities on global properties of world models \cite{buchert:dust},\cite{GBC}: in flat space, the relevant terms describing a nonvanishing impact are divergences of vector fields and as a result vanish for isolated systems and for periodic boundary conditions corresponding to a 3-torus topological architecture \cite{buchert:average}. For nonflat spatial sections, i.e. in the nonintegrable situation, backreaction terms in general furnish a global contribution.

The reader may find examples where the proposed correspondence has already been successfully employed in the transition from Lagrangian perturbation solutions in Newtonian theory to corresponding general-relativistic perturbation solutions, see the series of papers following \cite{rza1}
and the recent review paper \cite{Universe}. We also point the reader to the construction of exact solutions of general relativity from Newtonian solutions, see \cite{Kasai1995}, \cite{rza6} for Szekeres class II 
solutions with their corresponding Euclidean class \cite{Buchert1989AA}, that appear as subclasses of first-order Lagrangian perturbation solutions at a FLRW (Friedmann-Lema\^\i tre-Robertson-Walker) background; for Szekeres class I solutions see \cite{beyond}.\\

It is possible to extend the proposed strategy to more general spacetimes. In order to describe e.g. vortical flows and more general fluids, we have to consider tilted flows within a general ADM (Arnowitt-Deser-Misner) foliation of spacetime. 
For this purpose we have to diffeomorph the exact basis to obtain the general metric form (with lapse $N$ and shift $N^i$), below exemplified for a comoving description, where the coordinate velocity is set to zero.
The line element reads:
\beq
{^4}ds^2 = - \frac{N^2}{\gamma^2} dt^2  + 2N {\rm v}_i \,dt \,dX^i + s_{ij}\, dX^i dX^j \ ,
\eeq
with the covariant 3-velocity ${\rm v}^i = (N^i / N)$, the induced spatial metric components $s_{ij}$, and a 4-velocity that is tilted with respect to the hypersurface normal \cite{BMR}: 
\beq
u^\mu = \frac{\gamma}{N}(1, 0,0,0) \quad;\quad u_{\mu} = (-\frac{N}{\gamma} , \gamma {\rm v}_i) \ .
\eeq
In the general comoving setting the appearance of the Lorentz factor $\gamma$ resurrects the causality constant, while the appearence of the lapse $N$ makes the 
time deformation nonintegrable. Rendering the tilted 4-velocity Lagrangian, $u^\mu = (1,0,0,0)$, as our strategy demands, a \textit{proper-time foliation} $\tau = \int (N/\gamma) d t = const.$,  i.e.~$N=\gamma$,  investigated in \cite{BMR}, can be considered. A correspondence in proper-time foliation can thus be set up in the spirit of what has been said in this Letter.
Its realization is the subject of work in progress. 
In the tilted foliation, gravitomagnetic extensions of Newton's theory, as proposed by Heaviside \cite{heaviside} (see also the comments in \cite{bm}), will become relevant. \\

\noindent
{\bf Acknowledgments:} {\small This work is part of a project that 
has received funding from the European Research Council under the European Union's Horizon 2020 research and innovation program (grant agreement ERC advanced grant 740021-ARTHUS, PI: TB). 
Thanks to Hamed Barzegar, Henk van Elst, Asta Heinesen and Pierre Mourier
for valuable remarks on the manuscript, and to an anonymous referee for insightful and constructive suggestions.}
\newcommand\eprintarXiv[1]{\href{http://arXiv.org/abs/#1}{arXiv:#1}}

\end{document}